\documentclass[prl,twocolumn,aps,showpacs]{revtex4}
\usepackage{graphicx}
\usepackage{bm}

\begin{document}
\date{\today}
\title{Conserving {\em and\/} gapless approximations for the composite bosons 
in terms of the constituent fermions}
\author{G.C. Strinati and P. Pieri}
\affiliation{Dipartimento di Fisica, UdR INFM, Universit\`{a} di Camerino, 
I-62032 Camerino, Italy}
\date{\today}

\begin{abstract}
A long-standing problem with the many-body approximations for  
interacting condensed bosons has been the dichotomy
between the ``conserving'' and ``gapless'' approximations, which 
either obey the conservations laws or satisfy the
Hugenholtz-Pines condition for a gapless excitation spectrum, in the order.
It is here shown that such a dichotomy does not exist for a system of 
\emph{composite bosons\/}, which form as bound-fermion pairs in the
strong-coupling limit of the fermionic attraction.
By starting from the constituent fermions, for which conserving 
approximations can be constructed for any value of the mutual 
attraction according to the Baym-Kadanoff prescriptions, it is shown that 
these approximations also result in a gapless excitation spectrum 
for the boson-like propagators in the broken-symmetry phase.
This holds provided the corresponding equations for the 
fermionic single- and two-particle Green's functions are solved 
self-consistently.
\end{abstract}


\maketitle

Many-body decriptions of a system of interacting consensed bosons have long 
been known \cite{HM-65} to fall into either one of two classes of
approximation schemes, which are alternatively consistent with the 
conservation laws (conserving approximations) or with the absence of a gap 
in the elementary excitations spectrum (gapless approximations).
Having to choose between these two types of approximations constitutes a
shortcoming of the many-body theory
for condensed bosons, as one would rather like to deal with approximations 
which are 
conserving \emph{and\/} gapless at the same time.

These approximation schemes have been conceived for interacting-boson systems 
like helium, for which the internal fermionic structure
is immaterial, due to the large amount of energy required to excite the 
internal fermionic degrees of freedom compared with the energy scales 
of the experiments.

Recent experimental advances with ultracold trapped Fermi atoms have made it
 possible to produce systems of \emph{composite bosons\/} (dimers), 
whose binding energy is comparable with the energy and temperature involved in
the experiments~\cite{exp-gen-cb}.
The Bose-Einstein condensation of the dimers has also been 
detected~\cite{exp-cond-cb}.
In these systems, the internal fermionic structure is definitely relevant, as 
the binding energy of the dimers can be tuned across threshold via 
a Fano-Feshbach resonance~\cite{FF}, sweeping from bound to unbound fermions 
and viceversa.

For these systems, it appears appropriate to construct the dynamical 
propagators of the condensed composite bosons in terms of the constituent 
fermions, 
by following the progressive quenching of the fermionic degrees of freedom as 
the fermionic attraction is increased.
For the constituent fermions, it is known that conserving approximations can
be constructed for any value of the mutual attraction even in the
broken-symmetry (superfluid) phase, via the Baym-Kadanoff 
prescriptions \cite{BK-61,Baym-62} which require the self-consistent solution
of the equations for the single- and two-particle Green's functions.
In this way, conservation laws can be regarded to be fulfilled not only in 
terms of the constituent fermions but also in terms of the composite bosons,
when the fermionic attraction gets sufficiently strong that the fermionic 
degrees of freedom (internal to the composite bosons) are progressively
quenched.

The question is whether such conserving approximations (for the 
constituent fermions and therefore for the composite bosons) could also
result into a gapless excitation spectrum for the propagators of the composite
 bosons in the broken-symmetry phase, the gapless condition being
required on  general grounds by the occurrence of a Goldstone 
mode \cite{Rickayzen}.
[The case of a homogeneous system will be specifically considered when 
discussing the absence of a gap in the bosonic excitation spectrum 
\cite{Griffin}.]

Purpose of this paper is to show that a given fermionic conserving 
approximation also results in a gapless excitation spectrum for the boson-like
propagators.
The long-standing dichotomy between conserving and gapless approximations can
thus be apparently overcome, when the bosons themselves are treated 
at a more fundamental level in terms of the constituent fermions.

We begin by generalizing to the composite bosons the theorem of 
Hugenholtz-Pines for ordinary bosons, by following the treatment of 
Ref.~\onlinecite{HM-65} 
in terms of formally exact propagators (see also Ref.~\onlinecite{Rickayzen}).
Fermionic conserving approximations to select approximate propagators for the 
composite bosons will be considered later. 
To this end, we define a bosonic-like field operator
\begin{equation}
\Psi_{B}({\bf r}) = \int \! d\bm{ \rho} \, \phi(\rho)  
\psi_{\downarrow}({\bf r}-\bm{ \rho}/2) \psi_{\uparrow}({\bf r}+
\bm{ \rho}/2)
\label{definition-boson-fermion} 
\end{equation}
for any value of the fermionic coupling, where $\psi_{\sigma}({\bf r})$ is
 a fermionic field operator with spin $\sigma$.
When the fermionic attraction is sufficiently strong, on physical grounds the 
(real and normalized) function $\phi(\rho)$ can be taken 
as the bound solution of the associated two-body problem.
At weaker coupling, a precise choice of $\phi(\rho)$
is not required.
For instance, it could be taken as the solution of a generalized Cooper 
problem, whereby the Fermi energy is replaced by the coupling- (and 
temperature-) 
dependent fermionic chemical potential $\mu$.
To break the gauge symmetry, the bosonic order parameter $\alpha({\bf r}) =
 \langle  \Psi_{B}({\bf r}) \rangle_{\eta}$ is defined as the thermal 
average of the operator (\ref{definition-boson-fermion}) within the restricted
 ($\eta$) ensemble of Ref.~\onlinecite{HM-65}.
With the Nambu representation for the fermionic field operator 
$(\Psi_{1}({\bf r}) = \psi_{\uparrow}({\bf r}),
\Psi_{2}({\bf r}) = \psi_{\downarrow}^{\dagger}({\bf r}))$, this thermal
 average can then be expressed in terms of the anomalous
fermionic single-particle Green's function ${\cal G}_{12}$
\begin{equation}
\langle  \Psi_{2}^{\dagger}({\bf r}-\frac{\bm{\rho}}{2}) \Psi_{1}({\bf r}+
\frac{\bm{\rho}}{2}) \rangle_{\eta}  =
 {\cal G}_{12}({\bf r}+\frac{\bm{\rho}}{2},{\bf r}-
\frac{\bm{\rho}}{2};\tau=0^{-})\phantom{11} 
\label{G-12}
\end{equation}
with imaginary time $\tau$.

In what follows, it is convenient to consider a generalized 
fermionic single-particle Green's function:
\begin{equation}
{\cal G}(1,1') \, = \, - \,  
\frac{\langle  T_{\tau}[S \,\Psi(1) \, \Psi^{\dagger}(1')] \rangle}{\langle  T_{\tau}[S] \rangle}
\label{generalized-G}
\end{equation}
with the notation $1=({\bf r}_{1}, \tau_{1}, \ell_{1})$ in terms of
 the Nambu spinor component $\ell$.
Here, $T_{\tau}$ is the imaginary-time-ordering operator,
 $\langle \cdots\rangle$ is a thermal average taken with the system 
grand-canonical Hamiltonian $K = H - \mu N$, 
$\Psi(1) = \exp\{K \tau_{1}\} \Psi_{\ell_{1}} ({\bf r_{1}}) \exp\{- K 
\tau_{1}\}$, and the operator
$S =\exp\{- \int d11' \Psi^{\dagger}(1) U(1,1') \Psi(1') \}$  
contains the source term
\begin{equation}
U(1,1') \, = \, \left( \begin{array}{cc} 
U_{n}({\bf r_{1}},{\bf r_{1'}};\tau_{1}) &   
U_{s}({\bf r_{1'}},{\bf r_{1}};\tau_{1})^{*}   \\
U_{s}({\bf r_{1}},{\bf r_{1'}};\tau_{1}) & -
 U_{n}({\bf r_{1}},{\bf r_{1'}};\tau_{1})
\end{array} \right)\delta(\tau_{1} - \tau_{1'}^{+})
\label{S-definition}
\end{equation}
with a normal ($U_{n}$) and a superfluid ($U_{s}$) component.
[The normal component $U_{n}$ will not be needed in the following, while
 the superfluid component $U_{s}$ will be allowed to vanish at the end of the 
calculation.]
In the static case (when $U$ does not depend on the imaginary time), the
 generalized definition (\ref{generalized-G}) concides with the
ordinary definition [like Eq.(\ref{G-12})] within the $\eta-$ensemble.
With the definition (\ref{generalized-G}), the bosonic order parameter is 
generalized as follows:
\begin{equation}
\alpha({\bf r}) \, = \, \int \! d\bm{\rho} \, \phi(\rho) \, 
{\cal G}_{12}({\bf r}+\frac{\bm{\rho}}{2},\tau;{\bf r}-\frac{\bm{\rho}}{2},\tau^{+})  \,\, .
\label{alpha-gereralized}
\end{equation}

Suppose now that $U_{s}({\bf r},{\bf r'};\tau)$ is varied by a 
small uniform change of phase $\delta \Phi$, such that
$\delta U_{s}({\bf r},{\bf r'};\tau) \cong i \delta \Phi
U_{s}({\bf r},{\bf r'};\tau)$.
This change can be reabsorbed by a canonical tranformation of the fermionic 
field operators, so that the corresponding change
of the order parameter (\ref{alpha-gereralized}) is given by 
$\delta \alpha({\bf r}) = - i \delta \Phi \alpha({\bf r})$ 
to the leading order in $\delta \Phi$. 
The change $\delta \alpha({\bf r})$ can be calculated \emph{alternatively} via
 the definitions (\ref{alpha-gereralized}) and
(\ref{generalized-G}), by performing the functional derivative of 
(\ref{generalized-G}) with respect to a
variation of $U_{s}$. One obtains:
\begin{eqnarray}
&&\delta \alpha({\bf r}) =  - \, \int \! d\bm{\rho} \, \phi(\rho) \,  
                                     \int \! d{\bf r_2} \, d{\bf r_2'} \int \!
 d\tau_2 \nonumber\\
&& \times \left[ 
L({\bf r}+\frac{\bm{\rho}}{2},\tau,1;{\bf r_2},\tau_2,2;
{\bf r}-\frac{\bm{\rho}}{2},\tau^+,2;
{\bf r_2'},\tau_2^+,1)\right.\nonumber\\
&&\;\;\;\;\;\;\;\;\;\;\; \times \, \delta U_{s}({\bf r_2},{\bf r_2'};\tau_2)^*   
\nonumber \\                         
&& + L({\bf r}+
\frac{\bm{\rho}}{2},\tau,1;{\bf r_2},\tau_2,1;{\bf r}-
\frac{\bm{\rho}}{2},\tau^+,2;{\bf r_2'},\tau_2^+,2)\nonumber\\
&&\;\;\;\;\;\;\;\;\;\;\;\left.\times \, \delta U_{s}({\bf r_2'},{\bf r_2};\tau_2) 
\right] \,\, . \label{delta-alpha}    
\end{eqnarray}
Here, $L(1,2,1',2') = {\cal G}_{2}(1,2,1',2') - {\cal G}(1,1') {\cal G}(2,2')$ 
is the two-particle correlation function
expressed in terms of the generalized fermionic two-particle Green's 
function
\begin{equation}
{\cal G}_{2}(1,2,1',2') \, = \, \frac{\langle  T_{\tau}[S \, \Psi(1) \, 
\Psi(2) \, 
\Psi^{\dagger}(2') \, \Psi^{\dagger}(1')] \rangle}
{\langle  T_{\tau}[S]\rangle}   \,\, , 
\label{generalized-G-2}
\end{equation}
and is obtained via the functional derivative $L(1,2,1',2') = - 
\delta {\cal G}(1,1')/\delta U(2',2)$.

By a similar token, for the adjoint $\alpha({\bf r})^{*}$ of $\alpha({\bf r})$
 one obtains 
$\delta \alpha({\bf r})^{*} =  i \delta \Phi \alpha({\bf r})^{*}$, as well as
\begin{eqnarray}
&&\delta \alpha({\bf r})^{*} =  - \, \int \! d\bm{\rho} \, \phi(\rho) 
\,  \int \! d{\bf r_2} \, d{\bf r_2'} \int \!
 d\tau_2 \nonumber\\
&& \times \left[ 
L({\bf r}-\frac{\bm{\rho}}{2},\tau,2;{\bf r_2},\tau_2,2;
{\bf r}+\frac{\bm{\rho}}{2},\tau^+,1;
{\bf r_2'},\tau_2^+,1)\right.\nonumber\\
&&\;\;\;\;\;\;\;\;\;\;\; \times \, 
\delta U_{s}({\bf r_2},{\bf r_2'};\tau_2)^* \nonumber \\                       
&& + \, L({\bf r}-
\frac{\bm{\rho}}{2},\tau,2;{\bf r_2},\tau_2,1;{\bf r}+
\frac{{\bm\rho}}{2},\tau^+,1;{\bf r_2'},\tau_2^+,2) \nonumber\\
&&\;\;\;\;\;\;\;\;\;\;\;\left.\times \, 
\delta U_{s}({\bf r_2'},{\bf r_2};\tau_2) 
\right]  \label{delta-alpha-star}    
\end{eqnarray}
in the place of (\ref{delta-alpha}).
[The quantity $\alpha({\bf r})^{*}$ is defined like in
 Eq.(\ref{alpha-gereralized}) with ${\cal G}_{21}$ 
replacing ${\cal G}_{12}$, and coincides with the complex conjugate of 
$\alpha({\bf r})$ in the static case.]

The static and uniform limit of the above results can be considered at this 
point.
Accordingly, without loss of generality we let  $U_{s}({\bf r},{\bf r'};\tau) 
\rightarrow U_{s} \,\phi(|{\bf r}-{\bf r'}|)$ in both  
Eqs.~(\ref{delta-alpha}) and (\ref{delta-alpha-star}), where $U_{s}$ is a 
(complex) constant and $\phi({\bf r})$ the same function of
Eq.(\ref{definition-boson-fermion}).
We also introduce the Fourier representation:
\begin{eqnarray}
&& L(1,2,1',2') = \int \! \frac{d{\bf p}}{(2\pi)^{3}}  \frac{1}{\beta} 
\sum_{n}  
\int \! \frac{d{\bf p'}}{(2\pi)^{3}} \frac{1}{\beta} 
\sum_{n'} \label{L-F-T} \\
&& \times \int \! \frac{d{\bf q}}{(2\pi)^{3}}\frac{1}{\beta} 
\sum_{\nu} e^{i ({\bf p}+{\bf q}) \cdot {\bf r_1}} \,  e^{i {\bf p'} \cdot 
{\bf r_2}} \,
           e^{- i {\bf p} \cdot {\bf r_1'}} \, e^{- i ({\bf p'}+{\bf q}) 
\cdot {\bf r_2'}}  \nonumber \\
&& \times e^{- i(\omega_{n} + \Omega_{\nu})\tau_1} \, e^{- i \omega_{n'}
\tau_2} \,
           e^{i \omega_{n} \tau_1'} \, e^{i(\omega_{n'} + \Omega_{\nu})
\tau_2'} \,\,\,
           L^{\ell_1 \ell_2'}_{\ell_1' \ell_2} (p,p';q)    \,\,\, .
\nonumber 
\end{eqnarray}
Here, ${\bf p}$, ${\bf p'}$, and ${\bf q}$ are wave vectors, 
$\omega_{n}=(2n+1)\pi/\beta$ ($n$ integer) is a
fermionic Matsubara frequency ($\beta$ being the inverse temperature), 
$\Omega_{\nu}=2 \pi \nu/\beta$ ($\nu$ integer) a bosonic
Matsubara frequency, and   
$p=({\bf p},\omega_{n})$, $p'=({\bf p'},\omega_{n'})$, and 
$q=({\bf q},\Omega_{\nu})$ is a four-vector notation.
By straightforward manipulations of Eqs.~(\ref{delta-alpha}) and 
(\ref{delta-alpha-star}), and by recalling the identities
$\delta \alpha({\bf r}) = - i \delta \Phi \alpha({\bf r})$ and 
$\delta U_{s}({\bf r},{\bf r'};\tau) = 
i \delta \Phi U_{s}({\bf r},{\bf r'};\tau)$ (plus their adjoints), one ends 
up with the matrix equation
\begin{equation}
\left( \begin{array}{c} \alpha   \\ - \alpha \end{array} \right) = 
\left( \begin{array}{cc} 
G^{1 1}_{2 2} (q \rightarrow 0) &   G^{1 2}_{2 1} (q \rightarrow 0)  \\
G^{2 1}_{1 2} (q \rightarrow 0) &   G^{2 2}_{1 1} (q \rightarrow 0)
\end{array} \right) 
\left( \begin{array}{c} U_{s}  \\ - U_{s} \end{array} \right)
\label{alpha-vs-U}
\end{equation}
for $U$ and $\alpha$ real, with the notation
\begin{eqnarray}
G^{\ell_1 \ell_2'}_{\ell_1' \ell_2} (q) & = & - \int 
\! \frac{d{\bf p}}{(2\pi)^{3}}  \frac{1}{\beta} \sum_{n}  
e^{i \omega_{n} 0^+} 
\int \! \frac{d{\bf p'}}{(2\pi)^{3}} \frac{1}{\beta} 
\sum_{n'} \, e^{i \omega_{n'} 0^+}   \nonumber \\
& \times &  \phi({\bf p}+{\bf q}/2) \phi({\bf p'}+{\bf q}/2)  
L^{\ell_1 \ell_2'}_{\ell_1' \ell_2} (p,p';q)  \,.
\label{G-bosonic} 
\end{eqnarray}
It can be readily verified that, in the limit $U_{s} \rightarrow 0$, the 
definition (\ref{G-bosonic}) corresponds to the four possible (normal and 
anomalous) 
bosonic-like propagators which can be constructed with the operator 
(\ref{definition-boson-fermion}) and its adjoint.

Before letting $U_s \to 0$ in Eq.(\ref{alpha-vs-U}), it is convenient to 
introduce the inverse of the matrix on its right-hand side 
and write
\begin{equation}
\left( \begin{array}{cc} 
G^{1 1}_{2 2} (q \to 0) &   G^{1 2}_{2 1} (q \rightarrow 0)  \\
G^{2 1}_{1 2} (q \rightarrow 0) &   G^{2 2}_{1 1} (q \rightarrow 0)
\end{array} \right) = \frac{1}{AD - BC}     
\left( \begin{array}{cc} 
D &   - B  \\
- C &   A
\end{array} \right)\, .
\label{G-inverse}
\end{equation}
Matrix inversion of Eq.(\ref{alpha-vs-U}) then yields the  
conditions 
\begin{equation}
A - B = 0 \,\,\,\,\,\,\,\,\, , \,\,\,\,\,\,\,\,\,\, C - D = 0 
\,\,\,\,\,\,\,\,\, ,    \label{conditions}
\end{equation}
in order to have a finite value for the order parameter $\alpha$ in the limit 
$U_s \to 0$.
These conditions are not independent from each other, since one can prove 
on general ground from time-reversal invariance that $A=D$ and $B=C$. 
The denominator in Eq.(\ref{G-inverse}) then reduces to $A D - B C = (A - B) 
\, (A + B)$, and vanishes owing to (\ref{conditions}).
This implies that the bosonic-like propagators (\ref{G-bosonic}) are singular 
when $q \rightarrow 0$, irrespective of the value of the
fermionic coupling.
In the present context, the condition $A - B = 0$ corresponds to the 
Hugenholtz-Pines theorem for ordinary bosons~\cite{HP}.
It implies, in particular, that the \emph{composite bosons\/}, which form 
when the fermionic attraction is strong enough, have a 
\emph{gapless} spectrum.

All considerations made so far hold for the exact fermionic single- 
[Eq.(\ref{generalized-G})] and two-particle [Eq.(\ref{generalized-G-2})]
Green's functions.
The crucial point to derive the result (\ref{alpha-vs-U}) was that the single-
 and two-particle fermionic Green's functions are related to 
each other via a functional differentiation in the presence of the external 
potential $U$.
When dealing with fermions, it is standard practice to explore this relation 
further \cite{BK-61,Baym-62}, by exploiting the Dyson's equation with
the self-energy $\Sigma$:
\begin{equation}
- {\cal G}^{-1}(1,2) = \frac{\partial}{\partial \tau_1} \delta(1,2) 
+ M(1,2) + U(1,2) + \Sigma(1,2) \, .      
\label{Dyson-eq} 
\end{equation}
In this expression, 
$M(1,2) = \tau^{3}_{\ell_1 \ell_2} \, (h({\bf r_1}) - \mu) \, \delta({\bf r_1}
 - {\bf r_2}) \, \delta(\tau_1 - \tau_2)$ where
$\tau^{3}$ is a Pauli matrix and $h({\bf r})$ is the single-particle
 Hamiltonian (which, in general, includes an external static potential). 
The two-particle correlation function $L$ is correspondingly obtained as:
\begin{eqnarray}
& &- L(1,2,1',2') = \frac{\delta {\cal G}(1,1')}{\delta U(2',2)}\nonumber\\  
&&= -\int \! d34 \, {\cal G}(1,3)  
\frac{\delta {\cal G}^{-1}(3,4)}{\delta U(2',2)} {\cal G}(4,1')=  
{\cal G}(1,2') {\cal G}(2,1')\nonumber\\                            
&& + \int \! d3456 \, {\cal G}(1,3) 
{\cal G}(6,1') 
\frac{\delta \Sigma(3,6)}{\delta {\cal G}(4,5)} (-) L(4,2,5,2')  \, . 
 \label{Bethe-Salpeter-eq} 
\end{eqnarray}
It thus satisfies the Bethe-Salpeter equation, with kernel 
$\delta \Sigma / \delta {\cal G}$ related to the kernel 
$\Sigma$ of the Dyson's equation (\ref{Dyson-eq}).
The limit $U \rightarrow 0$ can be taken in Eqs.~(\ref{Dyson-eq}) 
and (\ref{Bethe-Salpeter-eq}) whenever appropriate.

Selection of an approximate fermionic many-body theory starts with an 
approximate choice of the functional form of the self-energy $\Sigma$ in 
terms of ${\cal G}$ (and of the two-body interaction).
The equations (\ref{Dyson-eq}) and (\ref{Bethe-Salpeter-eq}) are then solved 
self-consistently, with the respective approximate
kernels $\Sigma$ and $\delta \Sigma / \delta {\cal G}$.
In addition, Eq.(\ref{Bethe-Salpeter-eq}) implies that Eq.(\ref{delta-alpha}) 
(and its adjoint (\ref{delta-alpha-star})) holds even for the 
approximate theory, since $L$ (with the 
approximate kernel $\delta \Sigma / \delta {\cal G}$)
still represents the functional derivative of ${\cal G}$ with respect to $U$.
The alternative result $\delta \alpha({\bf r}) = 
- i \delta \Phi\alpha({\bf r})$ is instead obtained in the approximate theory
by noting that, under the transformation $U_{s}({\bf r},{\bf r'};\tau) 
\rightarrow e^{i \delta \Phi} U_{s}({\bf r},{\bf r'};\tau)$,
the approximate off-diagonal single-particle Green's function ${\cal G}_{12}$ 
of Eq.(\ref{alpha-gereralized}) trasforms as
${\cal G}_{12} \rightarrow e^{i \delta \Phi} {\cal G}_{12}$.
As a consequence, the result (\ref{conditions}) follows even within the 
approximate theory, implying a {\em gapless\/} spectrum.

\emph{Conserving} approximations for fermions are similarly based on 
Eqs.~(\ref{Dyson-eq}) and (\ref{Bethe-Salpeter-eq}), 
which are solved self-consistently for a given approximate choice of 
$\Sigma$.
In this case, the self-energy has to be chosen appropriately,  
to comply with the requirements of local number conservation
and gauge invariance \cite{Schrieffer}.
It is then required that the symmetry property $L(1,2,1',2')=L(2,1,2',1')$ is 
satisfied by the approximate $L$.
To this end, it is sufficient that the approximate kernel $\delta \Sigma / 
\delta {\cal G}$ of Eq.~(\ref{Bethe-Salpeter-eq}) 
satisfies the same symmetry property. 
This property is, in turn, met by any $\Phi-$derivable approximation for the
self-energy $\Sigma$ of Eq.(\ref{Dyson-eq}), whereby 
$\Sigma(1,2) = \delta \Phi / \delta {\cal G}(2,1)$ is obtained from
an approximate functional $\Phi$. \cite{Baym-62}
Attention must be paid to the fact that a given choice for 
$\Sigma$ may not 
meet this requirement, unless certain diagrams for $\Sigma$ are taken 
together \cite{Baym-62,RNC}.

It is now evident that a conserving approximation for the constituent 
fermions, which holds for any value of their mutual attraction, will also
result in a conserving approximation for the composite bosons that form in 
strong coupling.
The same fermionic conserving approximation will further result in a gapless 
spectrum for the composite bosons, as the same requirement for
Eqs.~(\ref{Dyson-eq}) and (\ref{Bethe-Salpeter-eq}) to be simultaneously 
self-consistently satisfied applies to both (conserving \emph{and}
gapless) procedures. 
This proves our claim.
From the above considerations, it is also clear that the requirements for a 
fermionic approximation to be conserving at any given coupling are 
more stringent than the absence of a gap in the 
excitation spectrum of the composite bosons in strong coupling.

\begin{figure}[ht]
\begin{center}
\includegraphics*[width=8.1cm]{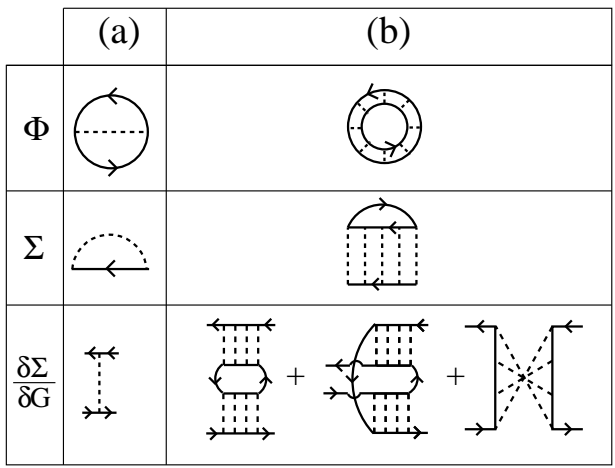}
\caption{Self-energy $\Sigma$ derived from the potential $\Phi$ with the 
associated kernel $\delta \Sigma / \delta {\cal G}$,
for (a) the BCS approximation and (b) the t-matrix 
approximation in the broken-symmetry phase.
Full lines represent fermionic (self-consistent) 
single-particle Green's functions, with the arrows pointing from the
second to the first argument; broken lines represent the 
fermionic interaction potential.}
\end{center}
\end{figure}

A well-known example of a fermionic conserving approximation, which results 
for any coupling in a gapless spectrum for the collective mode associated
with the broken symmetry \cite{Schrieffer}, is the BCS (off-diagonal) 
approximation for $\Sigma$, shown in Fig.1(a) with the associated potential
$\Phi$ and kernel $\delta \Sigma / \delta {\cal G}$ of the Bethe-Salpeter 
equation (\ref{Bethe-Salpeter-eq}).
In this case, the self-consistent solution of the Dyson's equation 
(\ref{Dyson-eq}) reduces to the solution of the BCS gap
equation; in addition, this equation coincides with the condition $A - B = 0$ 
of Eq.(\ref{conditions}) which 
guarantees the absence of a gap in the bosonic excitation spectrum.
In this simple case, therefore, the self-consistent solution of the BCS gap 
equation is sufficient for the approximation
to be conserving {\em and\/} gapless.
More generally, separate solutions of the equations (\ref{Dyson-eq}) and 
(\ref{Bethe-Salpeter-eq}) are required for the approximation to
be conserving and gapless.
For instance, to the self-energy $\Sigma$ within the fermionic t-matrix 
approximation in the broken-symmetry phase \cite{Haussmann}, shown in Fig.1(b) 
with the associated potential $\Phi$, there correspond three distinct 
contributions to the kernel $\delta \Sigma / \delta {\cal G}$, also shown in 
Fig.1(b).
When considering the BCS and t-matrix approximations for $\Sigma$ 
together, to get a gapless spectrum it is thus not enough to solve
self-consistently the Dyson's equation for ${\cal G}$ with both self-energy 
contributions, if one solves at the same time the Bethe-Salpeter 
equation with only the BCS contribution to kernel 
$\delta \Sigma / \delta {\cal G}$.
By doing so, one would, in fact, omit the three contributions to 
$\delta \Sigma / \delta {\cal G}$ depicted in Fig.1(b), whose presence is
required by conserving criteria.
Additional conserving \emph{and} gapless approximations can be similarly 
constructed by suitable choices of the fermionic self-energy.

In conclusion, we have shown that a given conserving approximation for the 
constituent fermions also results into a gapless spectrum for the 
composite bosons.
By following the formation of the bosons from the constituent fermions as the 
fermionic attraction is progressively increased, 
a long-standing (conserving vs gapless) dichotomy can
thus be resolved, at least at a formal level.
Although the self-consistent solution of the equations determining the 
fermionic single- and two-particle Green's functions might, in general, involve
considerable numerical labor, enforcing the 
fermionic conserving criteria proves \emph{per se} sufficient to get a 
gapless bosonic spectrum.

We are indebted to A. Fetter, F. Iachello, and G. Morandi for a critical 
reading of the manuscript. 
This work was supported by the Italian MIUR (contract 
Cofin-2003 ``Complex Systems and Many-Body Problems").

\end{document}